\documentclass[aps,prl,twocolumn,superscriptaddress]{revtex4-1}
\usepackage{breakurl}
\usepackage{xcolor}
\usepackage{sidecap}
\usepackage{amssymb}
\usepackage{hhline}
\usepackage{multirow}
\sidecaptionvpos{figure}{t}
\usepackage{amsmath}
\usepackage{mathrsfs}
\usepackage{graphicx}
\usepackage{esint}
\usepackage{epstopdf}
\usepackage{rotating}
\graphicspath{{pict/}{./}}
\usepackage{bm}%
\usepackage{microtype,bm,bbm,graphicx,booktabs,times}

\usepackage[normalem]{ulem}

\renewcommand {\Im}{\mathop\mathrm{Im}\nolimits}
\renewcommand {\Re}{\mathop\mathrm{Re}\nolimits}
\newcommand {\Tr}{\mathop\mathrm{Tr}\nolimits}
\renewcommand {\i}{{\rm i}}
\renewcommand {\phi}{{\varphi}}
\newcommand {\rmi}{{\rm i}}
\newcommand {\rmd}{{\rm d}}
\newcommand {\sign}{\mathop{\mathrm{sign}}\nolimits}
\newcommand {\e}{{\rm e}}
\newcommand {\eps}{\varepsilon}
\newcommand{\done}{\marginpar{\checkmark}}
\newcommand {\rot}{\mathop\mathrm{rot}\nolimits}

\newcounter{Fig}

\begin{document}

\title{Extremize Optical Chiralities through Polarization Singularities}
\author{Weijin Chen}
\email{Authors contribute equally to this work.}
\affiliation{School of Optical and Electronic Information, Huazhong University of Science and Technology, Wuhan, Hubei 430074, P. R. China}
\author{Qingdong Yang}
\email{Authors contribute equally to this work.}
\affiliation{School of Optical and Electronic Information, Huazhong University of Science and Technology, Wuhan, Hubei 430074, P. R. China}
\author{Yuntian Chen}
\email{Email: yuntian@hust.edu.cn}
\affiliation{School of Optical and Electronic Information, Huazhong University of Science and Technology, Wuhan, Hubei 430074, P. R. China}
\affiliation{Wuhan National Laboratory for Optoelectronics, Huazhong University of Science and Technology, Wuhan, Hubei 430074, P. R. China}
\author{Wei Liu}
\email{Email: wei.liu.pku@gmail.com}
\affiliation{College for Advanced Interdisciplinary Studies, National University of Defense
Technology, Changsha, Hunan 410073, P. R. China}
%

\begin{abstract}
 Chiral optical effects are generally quantified along some specific incident directions of exciting waves (especially for extrinsic chiralities of achiral structures) or defined as direction-independent properties by averaging the responses among all structure orientations. Though of great significance for various applications, chirality extremization (maximized or minimized) with respect to incident directions or structure orientations have not been explored, especially in a systematic manner. In this study we examine the chiral responses of open photonic structures from perspectives of quasi-normal modes and polarization singularities of their far-field radiations. The nontrivial topology of the  momentum sphere secures the existence of singularity directions along which mode radiations are either circularly or linearly polarized. When plane waves are incident along those directions, the reciprocity ensures ideal maximization and minimization of optical chiralities, for corresponding mode radiations of circular and linear polarizations respectively. For directions of general elliptical polarizations, we have unveiled the subtle equality of a Stokes parameter and the circular dichroism, showing that an intrinsically chiral structure can unexpectedly exhibit no chirality at all or even chiralities of opposite handedness for different incident directions. The framework we establish can be applied to not only finite scattering bodies but also infinite periodic structures, encompassing both intrinsic and extrinsic optical chiralities. We have effectively merged two vibrant disciplines of chiral and singular optics, which can potentially trigger more optical chirality-singularity related interdisciplinary studies.
\end{abstract}

\maketitle

Optical responses of photonic structures are generally anisotropic, depending on both polarizations and incident directions of exciting waves. Chiralities emerge when optical responses are distinct for circularly polarized waves of opposite handedness while the same incident direction~\cite{BARRON_2009__Molecular,BORISKINA__Singular}. Chiral effects can be observed for both intrinsically chiral and achiral structures~\cite{HENTSCHEL_2017_Sci.Adv._Chiral,COLLINS_AdvancedOpticalMaterials_chirality_2017,LUO_Adv.Opt.Mater._Plasmonic,QIU_Adv.Funct.Mater._3D,MUN_LightSci.Appl._Electromagnetic}. As for chiral structures, (intrinsic) chiralities are generally present and can be quantified as direction-independent properties through orientation-averaging chiral effects~\cite{BARRON_2009__Molecular,BORISKINA__Singular,FERNANDEZ-CORBATON_2015_ACSPhotonics_Dual,VINEGRAD_ACSPhotonics_Circular,SURYADHARMA_Phys.Rev.B_Quantifying,FAZEL-NAJAFABADI_ArXiv201210010Phys._Orientation};  while for achiral structures, (extrinsic) chiralities are manifest for some symmetry-breaking incident directions~\cite{PAPAKOSTAS_Phys.Rev.Lett._optical_2003,PLUM_Appl.Phys.Lett._Optical,PLUM_Phys.Rev.Lett._metamaterials_2009,SERSIC_Phys.Rev.Lett._Ubiquity,RIZZA_Phys.Rev.Lett._OneDimensional}, which nevertheless would be extinguished for symmetry-preserving directions or when orientation-averaged. For both scenarios, chiral responses are generally dependent on structure orientations, and thus it is interesting and significant to clarify systematically how chiralities can be extremized with respect to incident directions.

Besides chiral optics, another photonic branch of singular optics has also gained enormous momentum from two-decade explosive developments of nanophotonics, fertilizing many related disciplines~\cite{GBUR_2016__Singular, BERRY_2017__HalfCentury,DENNIS_2009_ProgressinOptics_Chapter,LIU_ArXiv201204919Phys._Topological}. Vectorial optical singularities correspond to states of circular and linear polarizations, for which   semi-major axes and orientation planes of polarization ellipses are undefined, respectively~\cite{NYE_1983_ProcRSocA_Polarization,NYE_1983_Proc.R.Soc.A_Lines,BERRY_2001_SecondInt.Conf.Singul.Opt.Opt.VorticesFundam.Appl._Geometry}. Polarization singularities are skeletons of general electromagnetic waves, which are robust against perturbations and generically manifest in both natural and artificial fields~\cite{GBUR_2016__Singular, BERRY_2017__HalfCentury,DENNIS_2009_ProgressinOptics_Chapter,LIU_ArXiv201204919Phys._Topological}.  Despite the ubiquity of polarization singularities, it is rather unfortunate and surprising that for decades both disciplines of chiral and singular optics develop almost independently with rare crucial interactions, especially when fundamental entities of circular polarizations are shared by both fields.

Here we investigate chirality extremization, for both achiral and chiral structures, with plane waves of variant incident directions. Chiral effects are examined from  perspectives of quasi-normal modes (QNMs) and their radiation polarization singularities. The nontrivial topology of momentum sphere has secured existences of singularity directions, as protected by the Poincar\'{e}-Hopf theorem~\cite{CHEN_2019__Singularities,CHEN_2019_ArXiv190409910Math-PhPhysicsphysics_Linea,CHEN_ACSOmega_Global}. For a structure supporting a dominant QNM at a spectral regime, with waves incident along singularity directions of circular and linear mode radiations,  chiral effects are ideally maximized and minimized respectively. For directions of general elliptical polarizations, the equality of a Stokes parameter and circular dichroism is discovered, demonstrating that an intrinsically chiral structure can unexpectedly manifest no chirality or even chiralities of opposite handedness for different orientations. The chirality extremization is generically protected by reciprocity and thus broadly applicable to both extrinsic and intrinsic chiral responses. 

Without loss of generality, we study  reciprocal nonmagnetic structures of relative permittivity $\mathbf{\epsilon }(\mathbf{r}, \omega)$  in vacuum background of refractive index $n=1$.  An open structure supports a set of QNMs characterized by eigenfield $\tilde{\mathbf{E}}_{j}(\mathbf{r})$ and complex eigenfrequency $\tilde{\omega}_{j}$~\cite{LALANNE__LaserPhotonicsRev._Light}. The incident plane wave is ${\mathbf{E}}_{\rm{inc}}(\mathbf{r})$, with wavevector  $\mathbf{k}_{\rm{inc}}$, real frequency $\omega$ and wavelength $\lambda$. Excited fields can be expanded into QNMs as $\mathbf{E}=\sum \alpha_{j}(\omega) \tilde{\mathbf{E}}_{j}(\mathbf{r})$, with  excitation coefficients~\cite{LALANNE__LaserPhotonicsRev._Light}:
\begin{equation}
\label{expansion-coefficient}
\alpha_{j}(\omega)\propto\iiint_{\mathbf{V}}-i \omega[\varepsilon(\mathbf{r}, \omega)-1] \mathbf{E}_{\rm{inc}}(\mathbf{r})\cdot \tilde{\mathbf{E}}_{j}(\mathbf{r}) d^{3} \mathbf{r},
\end{equation}
where $\mathbf{V}$ denotes regions the structure occupies. When only one non-degenerate QNM is dominantly excited, as is the case through this study, the mode subscript $j$ can be dropped. It is well-known that far-field radiations of QNMs and thus the corresponding scattered fields are divergent. Several approaches exist that can overcome this difficulty~\cite{LALANNE__LaserPhotonicsRev._Light}, and a direct technique is to treat $\tilde{\mathbf{J}}(\mathbf{r})=-i \omega[\varepsilon(\mathbf{r}, \omega)-1] \tilde{\mathbf{E}}(\mathbf{r})$ as source currents. They can then be expanded into electromagnetic multipoles (spherical harmonics) at real $\omega$, based on which convergent radiated (or scattered) far-fields $\tilde{\mathbf{E}}_{\rm{rad}}(\mathbf{r})$ can be directly calculated~\cite{POWELL_Phys.Rev.Applied_interference_2017,GRAHN_NewJ.Phys._electromagnetic_2012}.

This current-radiation perspective can simplify Eq.~(\ref{expansion-coefficient}) as:
\begin{equation}
\label{expansion-coefficient2}
\alpha(\omega)\propto\iiint_{\mathbf{V}}\mathbf{E}_{\rm{inc}}(\mathbf{r})\cdot\tilde{\mathbf{J}}(\mathbf{r}) d^{3} \mathbf{r}.
\end{equation}
Let's assume that ${\mathbf{E}}_{\rm{inc}}$ comes from a point-dipole moment $\mathscr{P}$ ($\mathscr{P}$ locates on the transverse plane $\mathscr{P} \cdot \mathbf{k}_{\rm{inc}}=0$ and is in-phase with ${\mathbf{E}}_{\rm{inc}}$: $\mathscr{P} \propto  {\mathbf{E}}_{\rm{inc}}$) in the far zone. According to Lorentz reciprocity~\cite{LANDAU_1984__Electrodynamicsb,POTTON_Rep.Prog.Phys._reciprocity_2004}:
\begin{equation}
\label{reciprocity-lorentz}
\iiint_{\mathbf{V}} \mathbf{E}_{\mathrm{inc}}(\mathbf{r}) \cdot \tilde{\mathbf{J}}(\mathbf{r}) d^{3} \mathbf{r}= \tilde{\mathbf{E}}_{\mathrm{rad}}\cdot \mathscr{\dot{P}}  =  -i \omega \tilde{\mathbf{E}}_{\mathrm{rad}} \cdot \mathscr{{P}},
\end{equation}
where time-derivative $\mathscr{\dot{P}}$ denotes the dipolar current and $\mathbf{E}_{\mathrm{rad}}$ is the radiated field along $\mathbf{k}_{\rm{rad}}=-\mathbf{k}_{\rm{inc}}$.  Since $\mathbf{E}_{\mathrm{inc}} \propto \mathscr{{P}}$, Eq.~(\ref{reciprocity-lorentz}) can be converted to $\iiint_{\mathbf{V}} \mathbf{E}_{\mathrm{inc}}(\mathbf{r}) \cdot \tilde{\mathbf{J}}(\mathbf{r}) d^{3} \mathbf{r}=-i \omega \tilde{\mathbf{E}}_{\mathrm{rad}} \cdot \mathscr{{P}}\propto \tilde{\mathbf{E}}_{\mathrm{rad}} \cdot \mathbf{E}_{\mathrm{inc}}$,
which simplifies Eq.~(\ref{expansion-coefficient2}) to:
\begin{equation}
\label{expansion-coefficient3}
\alpha(\omega)\propto \tilde{\mathbf{E}}_{\mathrm{rad}} \cdot \mathbf{E}_{\mathrm{inc}}, ~~\mathbf{k}_{\rm{inc}}=-\mathbf{k}_{\rm{rad}}.
\end{equation}
It means that QNM excitation efficiency for incident plane waves can be calculated directly, through the dot product of radiated and incident fields. Alternative derivations for Eq.~(\ref{expansion-coefficient3}), without referring to source currents of either radiated or incident fields, are presented in Ref.~\cite{Supplemental_Material}.

\begin{figure}[tp]
\centerline{\includegraphics[width=8.5cm]{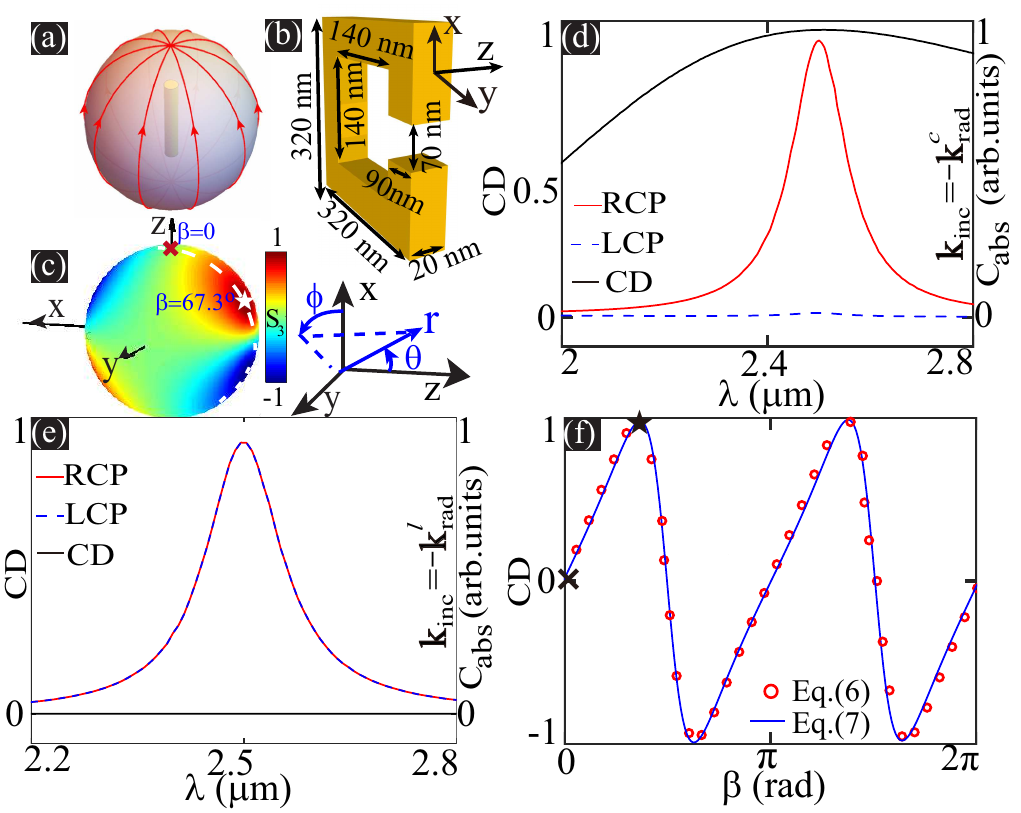}} \caption{\small  (a) A metal bar supports an electric dipole mode and its radiated electric fields on the momentum sphere. (b) A gold SRR with the spherical coordinate system $(r, \theta, \phi)$, and (c) $S_3$ distribution for the QNM supported. Two singularity directions are marked:  $\mathbf{k}_{\rm{rad}}^{c} (\theta=67.3^\circ, \phi=175.1^\circ)$ ($\star$) and $\mathbf{k}_{\rm{rad}}^{l} (\theta=0)$ ($\times$). (d) and (e) CD and absorption spectra for RCP and LCP incident waves, for $\mathbf{k}_{\rm{inc}}=-\mathbf{k}_{\rm{rad}}^{c}, -\mathbf{k}_{\rm{rad}}^{l}$ respectively [directions also marked in (f)]. (f) Angular CD spectra for waves incident antiparallel to directions on a great momentum circle marked in (c) by a dashed arc.}
\label{fig1}
\end{figure}

The validity of Eq.~(\ref{expansion-coefficient3}) can be checked against the simplest example of a metal bar supporting solely an electric dipole.  Electric fields radiated are shown in Fig.~\ref{fig1}(a): the dipole is vertically oriented and the fields are parallel to the lines of longitude~\cite{CHEN_ACSOmega_Global,JACKSON_1998__Classical}. Along the dipole, there are no radiations $\tilde{\mathbf{E}}_{\mathrm{rad}}=0$ and thus $\alpha=0$ according to Eq.~(\ref{expansion-coefficient3}). It means plane waves incident parallel to the bar will not excite it, which agrees with our conventional method of projecting the incident electric field onto the bar axis. Similar analysis can be conducted along other directions, where one linear polarization (parallel to the radiated field) excites the dipole while the other orthogonal polarization does not. The superiority of Eq.~(\ref{expansion-coefficient3}) to Eq.~(\ref{expansion-coefficient}) is not obvious for this example but would become apparent for sophisticated bodies.

Radiations can be always expanded into right- and left-handed circularly polarized (RCP and LCP, denoted respectively by $\circlearrowright$ and $\circlearrowleft$) components along $\mathbf{k}_{\rm{rad}}$: $\tilde{\mathbf{E}}_{\mathrm{rad}}=\mu\tilde{\mathbf{E}}_{\mathrm{rad}}^{\circlearrowright}+\nu\tilde{\mathbf{E}}_{\mathrm{rad}}^{\circlearrowleft}$. Here $\mu$ and $\nu$ are generally complex numbers ($|\mu^2|+|\nu^2|=1$), the ratio of which decides  properties of the polarization ellipse of $\tilde{\mathbf{E}}_{\mathrm{rad}}$~\cite{YARIV_2006__Photonics}. For incident RCP and LCP waves along $\mathbf{k}_{\rm{inc}}=-\mathbf{k}_{\rm{rad}}$, according to  Eq.~(\ref{expansion-coefficient3}) excitation coefficients are respectively:
\begin{equation}
\label{expansion-coefficient-circular}
\alpha^{\circlearrowright} \propto \mu, ~~ \alpha^{\circlearrowleft} \propto \nu.
\end{equation}
When only one QNM is dominantly excited, all cross sections of extinction,scattering and absorption are proportional to excitation efficiency $|\alpha|^2$: $\rm {C}_{\rm{ext,sca,abs}}^{\circlearrowright,\circlearrowleft} \propto |\alpha^{\circlearrowright,\circlearrowleft} |^2$. The circular dichroism (CD) is:
\begin{equation}
\label{CD-original}
\rm{CD}=(\rm {C}_{\rm{abs}}^{\circlearrowright}-\rm {C}_{\rm{abs}}^{\circlearrowleft})/(\rm {C}_{\rm{abs}}^{\circlearrowright}+\rm {C}_{\rm{abs}}^{\circlearrowleft}),
\end{equation}
which can then be simplified through Eq.~(\ref{expansion-coefficient-circular}) as:
\begin{equation}
\label{CD}
\rm{CD}=|\mu^2|-|\nu^2|=S_3,~~\mathbf{k}_{\rm{rad}}=-\mathbf{k}_{\rm{inc}}.
\end{equation}
Here $S_3$ is nothing but exactly one of the Stokes parameters widely employed for polarization characterizations~\cite{YARIV_2006__Photonics}: $S_3=\pm1$ corresponds respectively to RCP and LCP waves; $S_3=0$ corresponds to linear polarizations and other values of $S_3$ to elliptical polarizations. Equation~(\ref{CD}) constitutes the main result of this study, which reveals the subtle connection between $S_3$ of radiations and CD when only one QNM is dominantly excited: $|\rm{CD}|$ can be maximized and minimized antiparallel to the \textbf{C}-direction ($S_3=\pm1$) and \textbf{L}-direction ($S_3=0$) to ideal values of $|\rm{CD}|=1$ and $\rm{CD}=0$, respectively. We emphasize that those chirality extremization directions are exactly where generic polarization singularities are present~\cite{NYE_1983_ProcRSocA_Polarization,NYE_1983_Proc.R.Soc.A_Lines,BERRY_2001_SecondInt.Conf.Singul.Opt.Opt.VorticesFundam.Appl._Geometry}. The nontrivial topology of the momentum sphere secures the existence of those singularity directions~\cite{CHEN_2019__Singularities,CHEN_2019_ArXiv190409910Math-PhPhysicsphysics_Linea,CHEN_ACSOmega_Global}, which guarantees broad applicability of our model.

We can generalize definitions of CD to $\rm{CD_{ext,sca}}=(\rm {C}_{\rm{ext,sca}}^{\circlearrowright}-\rm {C}_{\rm{ext,sca}}^{\circlearrowleft})/(\rm {C}_{\rm{ext,sca}}^{\circlearrowright}+\rm {C}_{\rm{ext,sca}}^{\circlearrowleft})$, and identical to Eq.~(\ref{CD}) we have $\rm{CD_{ext,sca}}=S_3$~\cite{Supplemental_Material}. Based on this relation we can deduce polarization properties of mode radiations from general scattering properties of the scatterer. For example, it has been proved that for oppositely incident waves on a reciprocal scatterer~\cite{SOUNAS_Opt.Lett.OL_extinction_2014,CHEN_2020_Phys.Rev.Research_Scatteringa}: $\rm{CD_{ext}}(\mathbf{k}_{\rm{inc}})=\rm{CD_{ext}}(-\mathbf{k}_{\rm{inc}})$, which immediately implies: $S_3(\mathbf{k}_{\rm{rad}})=S_3(-\mathbf{k}_{\rm{rad}})$ and thus Eq.~(\ref{CD}) is actually valid for both directions of $\mathbf{k}_{\rm{inc}}=\pm\mathbf{k}_{\rm{rad}}$. It is further proved ~\cite{CHEN_2020_Phys.Rev.Research_Scatteringa} that for inversion-symmetric scatterers, $\rm{CD_{ext}}=0$ for arbitrary incident direction, which requires that QNM is everywhere linearly-polarized ($S_3=0$) throughout the momentum sphere~\cite{Supplemental_Material}.

It is clear from Eq.~(\ref{CD}) that CD (including its generalized versions of $\rm{CD_{ext,sca}}$) is solely decided by polarizations while has nothing to do with intensities of radiations. As a result, an alternative approach for expansions of the source currents $\tilde{\mathbf{J}}(\mathbf{r})$ into multipoles of complex frequencies does not affect our results, since such an expansion changes only the radiation intensity but not the polarization distributions (ratios between different multipoles are the same for real and complex frequency multipolar expansions)~\cite{WU_Phys.Rev.A_Intrinsic}.


To confirm our theory, we begin with the widely employed split ring resonator (SRR) that exhibits two symmetry mirrors  (\textbf{x-y} and \textbf{y-z} planes) and thus achiral [Fig.~\ref{fig1}(b)].  The SRR is made of gold with permittivity listed in Ref.~\cite{Johnson1972_PRB}. Neighbouring $\omega_1=7.4875\times10^{14}$ rad/s ($\lambda_1=2.516~\mu$m), an individual QNM is excited with eigenfrequency $\tilde{\omega}_{1}=(7.4875\times10^{14}-2.0414\times10^{13}\rm{i})$ rad/s (numerical results are obtained using COMSOL Multiphysics in this study).  The $S_3$ distribution for this mode is presented in Fig.~\ref{fig1}(c), where polarizations singularizes with $S_3=\pm 1$ or $0$ are generically manifest along various directions~\cite{GBUR_2016__Singular, BERRY_2017__HalfCentury,DENNIS_2009_ProgressinOptics_Chapter,LIU_ArXiv201204919Phys._Topological}. Two singularity directions are marked (one \textbf{C}-direction of $\mathbf{k}_{\rm{rad}}^{c}$ and one \textbf{L}-direction   of $\mathbf{k}_{\rm{rad}}^l\|\mathbf{z}^+$) and circularly-polarized plane waves are incident antiparallel to those directions. The corresponding spectra of absorption and CD are demonstrated respectively in Figs.~\ref{fig1}(d) and (e),  showing clearly that $|\rm{CD}|$ is maximized and minimized respectively. The deviation of CD from its ideal absolute values of $1$ and $0$ at some spectral positions is induced by some marginal contributions of other QNMs excited, as is also the case in following studies of other structures. We emphasize that zero CD in Fig.~\ref{fig1}(e) is observed along $\mathbf{k}_{\rm{rad}}^l$ that preserves the mirror symmetry of the whole scattering configuration (parity conservation requires zero CD~\cite{BARRON_2009__Molecular,CHEN_2020_Phys.Rev.Research_Scatteringa}). This is consistent with the conception that extrinsic chiralities are present along symmetry-breaking directions~\cite{PAPAKOSTAS_Phys.Rev.Lett._optical_2003,PLUM_Appl.Phys.Lett._Optical,PLUM_Phys.Rev.Lett._metamaterials_2009,SERSIC_Phys.Rev.Lett._Ubiquity,RIZZA_Phys.Rev.Lett._OneDimensional}.

To verify the validity of  Eq.~(\ref{CD}) for general non-singularity directions $0<|S_3|<1$, a great circle on the momentum sphere (parameterized by $0\leq\beta\leq2\pi$) that contains the marked \textbf{C}-direction and \textbf{L}-direction is selected [also marked in Fig.~\ref{fig1}(c)]. The angular CD spectra ($\omega=\omega_1$) along directions antiparallel to points on this circle are shown in Fig.~\ref{fig1}(f), where two sets of results are presented: one calculated from $S_3$ [Eq.~(\ref{CD})] and the other through $\rm {C}_{\rm{abs}}$  obtained in direct scattering simulations [Eq.~(\ref{CD-original})]. Both sets of results agree well, the symmetry of which [$\rm{CD}(\beta)\approx\rm{CD}(\beta\pm\pi)$] reconfirms our previous claims of $\rm{CD}(\mathbf{k}_{\rm{inc}})=\rm{CD}(-\mathbf{k}_{\rm{inc}})$ and $S_3(\mathbf{k}_{\rm{rad}})=S_3(-\mathbf{k}_{\rm{rad}})$ for single mode excitations.

Now that achiral structures do exhibit chiral responses along some incident directions, why should they be classified as achiral?  The conventional standard is purely geometric: they exhibit mirror or inversion symmetries and thus can be superimposed onto their mirror images~\cite{BARRON_2009__Molecular}. Alternatively, CD can be employed to categorize structures as chiral or not, but only when it is orientation-averaged among all incident directions:
\begin{equation}
\label{CD-average}
\mathbb{CD}=\frac{1}{4\pi |\mathbf{k}_{\rm{inc}}|^2}\iint\rm{CD}~d^{2}\mathbf{k}_{\rm{inc}}.
\end{equation}
The law of parity conservation~\cite{BARRON_2009__Molecular,CHEN_2020_Phys.Rev.Research_Scatteringa} ensures that the mirror-symmetry or inversion-symmetry operation maps $\rm {C}_{\rm{abs}}^{\circlearrowright,\circlearrowleft}$ to $\rm {C}_{\rm{abs}}^{\circlearrowleft,\circlearrowright}$, flipping the sign of CD according to Eq.~(\ref{CD-original}). As a result, $\mathbb{CD}=0$ for mirror-symmetric or inversion-symmetric structures, which agrees with the geometric standard of image superimposition.

%

%
\begin{figure}[tp]
\centerline{\includegraphics[width=8.5cm]{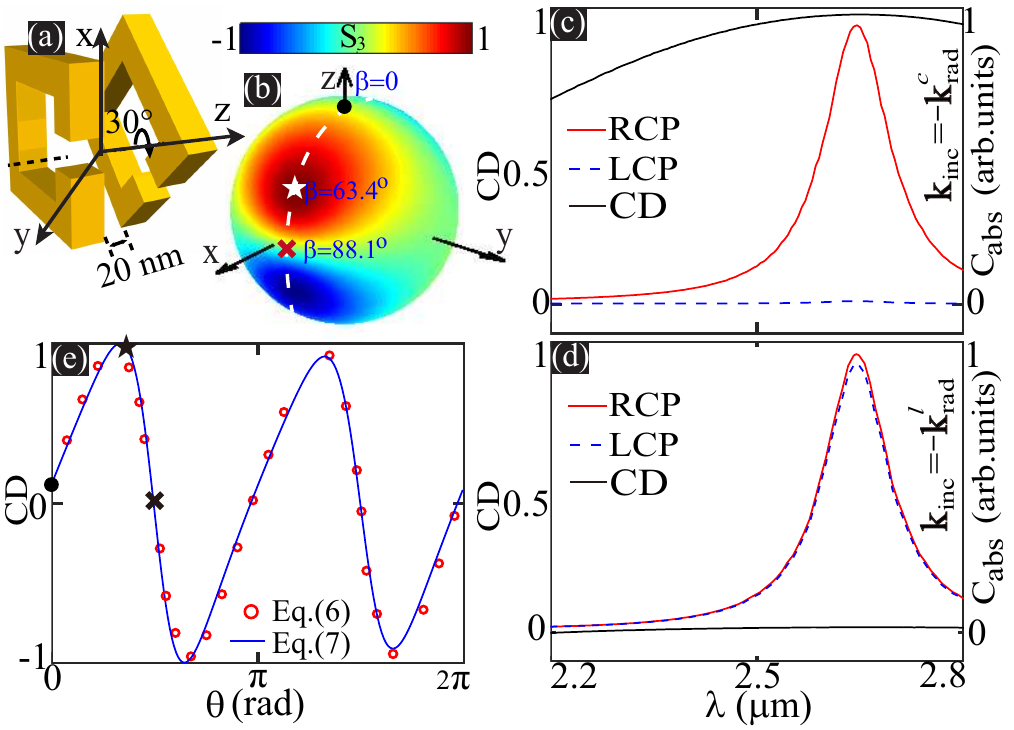}} \caption{\small (a) An intrinsically chiral SRR dimer and (b) $S_3$ distribution for the QNM supported. Two singularity directions are marked:  $\mathbf{k}_{\rm{rad}}^{c} (\theta=63.4^\circ, \phi=7.3^\circ)$ ($\star$) and $\mathbf{k}_{\rm{rad}}^{l} (\theta=88.1^\circ, \phi=7.3^\circ)$ ($\times$). (c) and (d) CD and absorption spectra for circularly-polarized incident waves), for $\mathbf{k}_{\rm{inc}}=-\mathbf{k}_{\rm{rad}}^{c}, -\mathbf{k}_{\rm{rad}}^{l}$ respectively [directions also marked in (e)]. (e) Angular CD spectra for waves incident antiparallel to directions on a great momentum circle marked in (b) by a dashed arc.}\label{fig2}
\end{figure}

Now we turn to a SRR dimer that is intrinsically chiral [Fig.~\ref{fig2}(a)], which supports a QNM at eigenfrequency $\tilde{\omega}_{2}=(7.0765\times10^{14}-1.6125\times10^{13}\rm{i})$ rad/s.  The $S_3$ distribution of this QNM is presented in Fig.~\ref{fig2}(b), where one \textbf{C}-direction and one \textbf{L}-direction are marked. The corresponding spectra of absorption and CD are demonstrated respectively in Figs.~\ref{fig2}(c) and (d) for plane waves incident antiparallel to those directions. We have also selected a great momentum circle [as marked in Fig.~\ref{fig2}(b), containing \textbf{C}-direction, \textbf{L}-direction and the \textbf{z}-axis] and the corresponding angular $\rm{CD}$ spectra [$\omega=\rm{Re}(\tilde{\omega}_{2})$ and $\lambda=~2.662~\mu$m, where Re denotes the real part] along directions antiparallel to points on this circle are shown in Fig.~\ref{fig2}(e).
It is worth noting that along the dimer twisting direction (\textbf{z}-axis with $\beta=0$ and $\pi$), CD is actually quite small ($\rm{CD}=S_3=0.12$) and far from being maximal.

For chiral structures consisting of twisted elements that are similar to that shown in Fig.~\ref{fig2}(a), though not stated explicitly, it is widely believed that the incident direction of maximal $|\rm{CD}|$ should be parallel to the twisting axis~\cite{BORISKINA__Singular,HENTSCHEL_2017_Sci.Adv._Chiral,COLLINS_AdvancedOpticalMaterials_chirality_2017,QIU_Adv.Funct.Mater._3D,MUN_LightSci.Appl._Electromagnetic,GANSEL_Science_Gold,SCHAFERLING_2012_Phys.Rev.X_Tailoring,YIN_NanoLett._Interpreting,CUI_NanoLett._Giant,WANG_ACSPhotonics_Circular,NAJAFABADI_2017_ACSPhotonics_Analytical,GORKUNOV_Phys.Rev.Lett._Metasurfaces}.
In fact, this is not the case and CD along other directions can be much more significant as shown in Fig.~\ref{fig2}(e).  A structure being achiral or chiral is decided by whether or not $\mathbb{CD}=0$, and the sign of $\mathbb{CD}$ rather than CD determines overall structure handedness. For achiral structures, $\mathbb{CD}=0$ does not require $\rm{CD}=0$ everywhere, and extrinsic chiralities emerge along directions of $\rm{CD}\neq0$ ~\cite{PAPAKOSTAS_Phys.Rev.Lett._optical_2003,PLUM_Appl.Phys.Lett._Optical,PLUM_Phys.Rev.Lett._metamaterials_2009,SERSIC_Phys.Rev.Lett._Ubiquity,RIZZA_Phys.Rev.Lett._OneDimensional}. In a similar fashion, for chiral structures, $\mathbb{CD}\neq0$ requires neither $\rm{CD}\neq0$ nor $\mathbb{CD}$ and $\rm{CD}$ being of the same sign everywhere. This means that an intrinsically chiral structure can manifest no chirality at all ($\rm{CD}=0$) or even chiralities of opposite handedness (CD of different signs) among different incident directions [Fig.~\ref{fig2}(e)].


Up to now, we have discussed how to extremize CD to its ideal absolute maximum or minimum values with respect to incident directions. Since $|\mathbb{CD}|$ is automatically minimized ($\mathbb{CD}=0$) for achiral structures,  it is both interesting and significant to ask how to maximize $|\mathbb{CD}|$ ideally to $\mathbb{CD}=\pm1$?  According to Eq.~(\ref{CD-average}), the ideal maximization of $|\mathbb{CD}|$ requires $\rm{CD}=\pm 1$ along all incident directions [or more accurately some isolated directions can be exempted, as they do not affect the overall integration in Eq.~(\ref{CD-average})].  This requires mode radiations are circularly-polarized with the same handedness throughout the momentum sphere except for some isolated directions.  This is possible, for example, for a pair of parallel electric and magnetic dipoles (or higher-order multipoles of the same order) of the same magnitude (in terms of total scattered power) and $\pm \pi/2$ phase contrast, as secured by the electromagnetic duality symmetry (along the isolated directions parallel to the dipole orientation direction there are no radiations and thus the polarizations are not defined)~\cite{JACKSON_1998__Classical}. Our conclusions for $|\mathbb{CD}|$ maximization are consistent with the results presented in Ref.~\cite{FERNANDEZ-CORBATON_2016_Phys.Rev.X_Objects}, where the problem has been approached from a very different perspective. However up to now, most if not all realistic reciprocal structures are anisotropic in terms of CD responses and thus ideal $\mathbb{CD}=\pm1$ has not been obtained so far, though not theoretically impossible.

\begin{figure}[tp]
\centerline{\includegraphics[width=9cm]{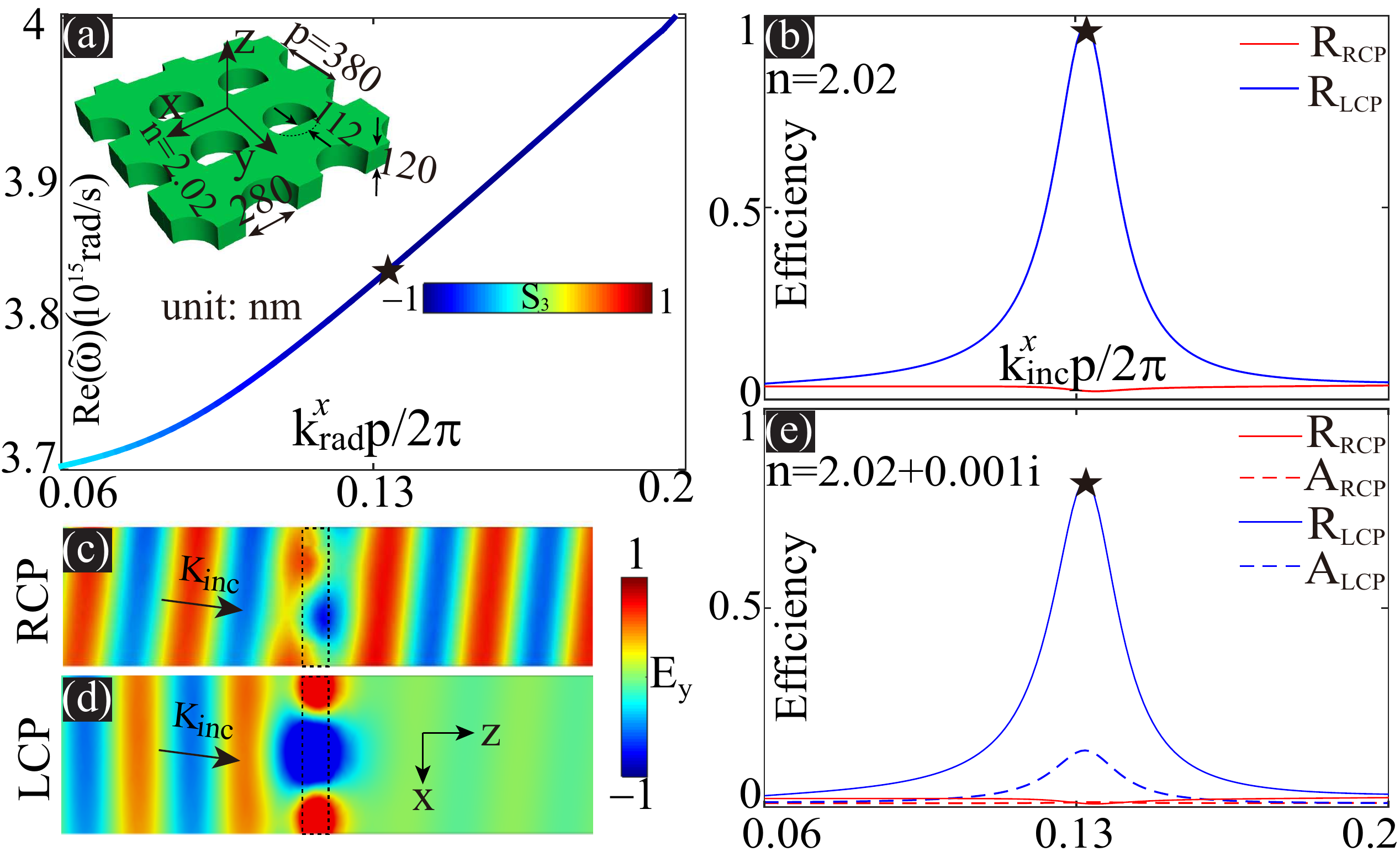}} \caption{\small (a) The dispersion curve of one TE-like QNM band for square-latticed PCS (show schematically as inset). The $S_3$ distribution of the QNM band is also shown and a singularity direction $\mathbf{k}_{\rm{rad}}^{c}$ is marked ($\star$). (b) Angular reflection spectra for RCP and LCP incident waves without material losses, and (c-d) corresponding near-field distributions when $\mathbf{k}_{\rm{inc}}=\mathbf{k}_{\rm{rad}}^c$.  (e) Angular absorption and reflection spectra with material losses.}\label{fig3}
\end{figure}

As a final step, we proceed to apply our framework to infinitely-extended photonic crystal slabs (PCSs), which support Bloch QNMs~\cite{JOANNOPOULOS_2008__Photonic,HSU_Nat.Rev.Mater._bound_2016,LIU_ArXiv201204919Phys._Topological}. It is well known that through symmetry breaking, some bound states in the continuum~\cite{HSU_Nat.Rev.Mater._bound_2016} can be broken into circularly-polarized radiating states~\cite{LIU_ArXiv201204919Phys._Topological,LIU_2019_Phys.Rev.Lett._Circularly,GUO_2020_Phys.Rev.Lett._Meron,YIN_2020_Nature_Observationa,YE_2020_Phys.Rev.Lett._Singular,CHEN_2019_ArXiv190409910Math-PhPhysicsphysics_Linea,YODA_Phys.Rev.Lett._Generationa}. One such square-latticed (periodicity $p=380$~nm) PCS of index $n=2.02$ is shown as inset of Fig.~\ref{fig3}(a), which is achiral (symmetric with respect to the \textbf{x}-\textbf{y} and \textbf{y}-\textbf{z} planes) while the symmetry respect to the \textbf{x}-\textbf{z} plane is broken. The dispersion curve [Re($\tilde{\omega}$) versus $\mathbf{k}_{\rm{rad}}^x$, with $\mathbf{k}_{\rm{rad}}^y=0$] that is colored according to the $S_3$ distribution of QNM radiations for one non-degenerate TE-like (electric fields distributed on the \textbf{x}-\textbf{y} plane) band are shown in Fig.~\ref{fig3}(a). One \textbf{C}-direction is marked ($\star$) with $\tilde{\omega}_{3}=(3.8409\times10^{15}-1.8686\times10^{13}\rm{i})$ rad/s and $\mathbf{k}_{\rm{rad}}^{c} =(\mathbf{k}_{\rm{rad}}^{c,x}, \mathbf{k}_{\rm{rad}}^{c,y}, \mathbf{k}_{\rm{rad}}^{c,z})$, with $\mathbf{k}_{\rm{rad}}^{c,x}p/2\pi=0.1332$, $\mathbf{k}_{\rm{rad}}^{c,y}=0$, $\mathbf{k}_{\rm{rad}}^{c,z}=\sqrt{[\rm{Re}(\tilde{\omega}_3)/c]^2-(\mathbf{k}_{\rm{rad}}^{c,x})^2-(\mathbf{k}_{\rm{rad}}^{c,y})^2}$ and $c$ is the speed of light.  Now we shine circularly-polarized plane waves onto the PCS with fixed frequency $\omega$=Re$(\tilde{\omega}_3)$ along different directions (variant $\mathbf{k}_{\rm{inc}}^x$) on the \textbf{x}-\textbf{z} plane ($\mathbf{k}_{\rm{inc}}^y=0$), with $\mathbf{k}_{\rm{inc}}=(\mathbf{k}_{\rm{inc}}^x, \mathbf{k}_{\rm{inc}}^y, \mathbf{k}_{\rm{inc}}^z)$.  
Angular reflection (R) spectra is presented in Fig.~\ref{fig3}(b), where the critical direction $\mathbf{k}_{\rm{inc}}=\mathbf{k}_{\rm{rad}}^c$ is marked.

Figure~\ref{fig3}(b) shows clearly that along \textbf{C}-direction RCP and LCP waves are almost fully reflected and transmitted (no higher-order diffractions), respectively. Near-field distributions ($\mathbf{E}_y$ on the \textbf{x}-\textbf{z} plane; dashed rectangles denote the cross sections of the PCS) are presented in Figs.~\ref{fig3}(c) and (d), where the formation of standing wave upon reflection in Fig.~\ref{fig3}(d) obscures the information of oblique incidence. Those properties agree with our result in Eq.~(\ref{expansion-coefficient3}), which requires that circularly-polarized Bloch QNM is maximally and not excited by LCP and RCP waves, leading to perfect reflection and transmission respectively. When material losses are incorporated ($n=2.02+0.001i$), both spectra of reflection and absorption (A) efficiency are summarized in Fig.~\ref{fig3}(e).  Along the marked critical direction, as expected, there is
considerable and almost no absorption for incident LCP and RCP waves respectively, which is also consistent with our previous analyses.



To conclude, we revisit optical chiralities from perspectives of QNMs and their polarization singularities. When plane waves are incident parallel to singularity \textbf{C}- and \textbf{L}-directions, the chiral responses in terms of CD would be ideally maximized and minimized, respectively. For general incident directions, we discover the subtle equivalence of CD and $S_3$ of the QNM radiation, showing that an intrinsical chiral structures can surprisingly manifest no chirality at all or even chiralities of opposite handedness among different incident directions. The validity of our conclusions resides on the approximation of single QNM excitation, which is indeed applicable to many structures at various spectral regimes, especially to the chiral ones that do not exhibit any geometric symmetries (\textit{e.g.} rotational symmetries that protect at least one pair of degenerate counter-rotating QNMs~\cite{GORKUNOV_Phys.Rev.Lett._Metasurfaces}).  When several QNMs (including degenerate ones) are co-excited, observable optical properties (such as $\rm {C}_{\rm{ext,sca, abs}}$) are neither proportional to the excitation efficiency of any QNM nor their direct sum, since QNMs are generally not orthogonal and inter-mode interference terms have to be carefully analysed~\cite{LALANNE__LaserPhotonicsRev._Light,POWELL_Phys.Rev.Applied_interference_2017,LEUNG_1994_Phys.Rev.A_Completeness}. Besides QNM-based approaches~\cite{LALANNE__LaserPhotonicsRev._Light,POWELL_Phys.Rev.Applied_interference_2017,ALPEGGIANI_Phys.Rev.X_quasinormalmode_2017-1}, techniques such as the scattering matrix method~\cite{MISHCHENKO_2002__Scattering} can be also employed to tackle more general scenarios, for which to extract simple and general principles like the ones unveiled here constitutes a promising future research direction.  Our results are directly applicable to elliptical dichroism~\cite{Supplemental_Material}, and singularity based approaches we introduce can be potentially extended to structured non-plane beams, especially those carrying both spin and angular momentum~\cite{ZAMBRANA-PUYALTO_Nat.Commun._Angular,NI_PNAS_Gigantic}. 

\emph{Acknowledgment}: We acknowledge the financial support from National Natural Science Foundation of China (Grant No. 11874026 and 11874426), and several other Researcher Schemes of National University of Defense Technology. W. L. is indebted to Sir Michael Berry and Prof. Tristan Needham for invaluable correspondences.

\onecolumngrid
\clearpage

\renewcommand {\Im}{\mathop\mathrm{Im}\nolimits}
\renewcommand {\Re}{\mathop\mathrm{Re}\nolimits}
\renewcommand {\i}{{\rm i}}
\renewcommand {\phi}{{\varphi}}
\renewcommand {\rmi}{{\rm i}}
\renewcommand {\rmd}{{\rm d}}
\renewcommand {\sign}{\mathop{\mathrm{sign}}\nolimits}
\renewcommand {\e}{{\rm e}}
\renewcommand {\eps}{\varepsilon}
\renewcommand{\done}{\marginpar{\checkmark}}
\renewcommand {\rot}{\mathop\mathrm{rot}\nolimits}
\renewcommand {\Tr}{{\rm Tr}\,}

\begin{center}
\noindent\textbf{\large{Supplemental Material for:}}
\\\bigskip
\noindent\textbf{\large{Extremize Optical Chiralities through Polarization Singularities}}
\\\bigskip
\onecolumngrid

Weijin Chen$^{1,*}$, Qingdong Yang$^{1,*}$, Yuntian Chen$^{1,2,\dag}$, and Wei Liu$^{3,\ddagger}$

\small{$^1$ \emph{School of Optical and Electronic Information, Huazhong University of Science and Technology, Wuhan, Hubei 430074, P. R. China}}\\
\small{$^2$ \emph{Wuhan National Laboratory for Optoelectronics, Huazhong University of Science and Technology, Wuhan, Hubei 430074, P. R. China}}\\
\small{$^3$ \emph{College for Advanced Interdisciplinary Studies, National University of Defense
Technology, Changsha, Hunan 410073, P. R. China}}
\end{center}

The Supplemental Material includes the following three sections: (\textbf{\uppercase\expandafter{\romannumeral1}}). Alternative derivations for Eq.~(4) without referring to source currents of radiations or incident fields;
(\textbf{\uppercase\expandafter{\romannumeral2}}). Verifications of $\rm{CD_{ext,sca}}=S_3$ for both achiral and chiral scatterers;
(\textbf{\uppercase\expandafter{\romannumeral3}}). $S_3$ distributions of  an inversion-symmetric scatterer and its CD responses; (\textbf{\uppercase\expandafter{\romannumeral4}}). Applications for elliptical dichroism maximization.

%


\setcounter{equation}{0}
\setcounter{figure}{0}
\newcounter{sfigure}
\setcounter{sfigure}{1}
\setcounter{table}{0}
\renewcommand{\theequation}{S\arabic{equation}}

 \renewcommand\thefigure{S{\arabic{figure}}}
\renewcommand{\thesection}{S\arabic{section}}

\renewcommand*{\citenumfont}[1]{S#1}
\renewcommand*{\bibnumfmt}[1]{[S#1]}

\section{(\textbf{\uppercase\expandafter{\romannumeral1}}).  Alternative derivations for Eq.~(4) without referring to source currents of radiations or incident fields}

According to de Hoop's formulations~\cite{DEHOOP_Appl.sci.Res._reciprocitya}, the direct dot product of $\tilde{\mathbf{E}}_{\mathrm{rad}}$ and $\mathbf{E}_{\mathrm{inc}}$ in the far zone can be expressed as an integral form (Eq. (4.10) of Ref.~\cite{DEHOOP_Appl.sci.Res._reciprocitya}) :

\begin{equation}
\label{hoop1}
\tilde{\mathbf{E}}_{\mathrm{rad}} \cdot \mathbf{E}_{\mathrm{inc}}\propto\int_{\mathbf{S}}\left({\mathbf{E}}^{\mathrm{s}}\times \mathrm{\mathbf{H}}_{\mathrm{inc}}^s-\mathrm{\mathbf{E}}_{\mathrm{inc}}^s \times {\mathbf{H}}^{\mathrm{s}} \right) \cdot \boldsymbol{n} \mathrm{d}\mathbf{S},  ~~\mathbf{k}_{\rm{inc}}=-\mathbf{k}_{\rm{rad}}.
\end{equation}
Here \textbf{S} denote the boundary of the scatterer and $\boldsymbol{n}$ denotes its normal vector pointing outwards; (${\mathbf{E}}^{\mathrm{s}}$,~${\mathbf{H}}^{\mathrm{s}}$) are radiated electric and magnetic fields on S of the QNM excited; ($\mathbf{E}_{\mathrm{inc}}^s$,~$\mathbf{H}_{\mathrm{inc}}^s$) are incident electric and magnetic fields on the boundary. Apparently (${\mathbf{E}}^{\mathrm{s}}$,~${\mathbf{H}}^{\mathrm{s}}$) originate from the excited QNM and thus for single-mode excitation we have:
\begin{equation}
\label{hoop2}
({\mathbf{E}}^{\mathrm{s}},{\mathbf{H}}^{\mathrm{s}})\propto \alpha(\tilde{{\mathbf{E}}},{\tilde{\mathbf{H}}}),
\end{equation}
where $(\tilde{{\mathbf{E}}},~{\tilde{\mathbf{H}}})$ are the eigenfields of the QNM excited. Through Eqs.~(\ref{hoop1}) and (\ref{hoop2}) we obtain:
\begin{equation}
\label{hoop4}
\tilde{\mathbf{E}}_{\mathrm{rad}} \cdot \mathbf{E}_{\mathrm{inc}}\propto\alpha\int_{\mathbf{S}}\left(\tilde{{\mathbf{E}}}\times \mathrm{\mathbf{H}}_{\mathrm{inc}}^s-\mathrm{\mathbf{E}}_{\mathrm{inc}}^s \times {\tilde{\mathbf{H}}} \right) \cdot \boldsymbol{n} \mathrm{d}\mathbf{S},
\end{equation}
which can be simplified as:
\begin{equation}
\label{hoop3}
\alpha\propto \tilde{\mathbf{E}}_{\mathrm{rad}} \cdot \mathbf{E}_{\mathrm{inc}}, ~~\mathbf{k}_{\rm{inc}}=-\mathbf{k}_{\rm{rad}}.
\end{equation}
It is exactly Eq. (4) in the main text.

We further note that the field-field surface integral form on the right hand side of  Eq.~(\ref{hoop1}) can be converted to the field-current volume integral form through the Lorentz reciprocity~\cite{LANDAU_1984__Electrodynamicsb}:
\begin{equation}
\label{hoop5}
\int_{\mathbf{S}}\left({\mathbf{E}}^{\mathrm{s}}\times \mathrm{\mathbf{H}}_{\mathrm{inc}}^s-\mathrm{\mathbf{E}}_{\mathrm{inc}}^s \times {\mathbf{H}}^{\mathrm{s}} \right) \cdot \boldsymbol{n} \mathrm{d}\mathbf{S}=\iiint_{\mathbf{V}}\mathbf{E}_{\rm{inc}}(\mathbf{r})\cdot\tilde{\mathbf{J}}(\mathbf{r}) d^{3} \mathbf{r},  ~~\mathbf{k}_{\rm{inc}}=-\mathbf{k}_{\rm{rad}},
\end{equation}
which is consist with Eqs. (2-4) in our main text.  As a result, the derivations formulated through Eqs.~(\ref{hoop1})-(\ref{hoop3}) are effectively equivalent to those in the main text, despite that here the source currents of neither the radiated nor the incident fields are involved.

\section{(\textbf{\uppercase\expandafter{\romannumeral2}}). Verifications of $\rm{CD_{ext,sca}}=S_3$ for both achiral and chiral scatterers}
\begin{figure}[htbp]
\centerline{\includegraphics[width=8.5cm]{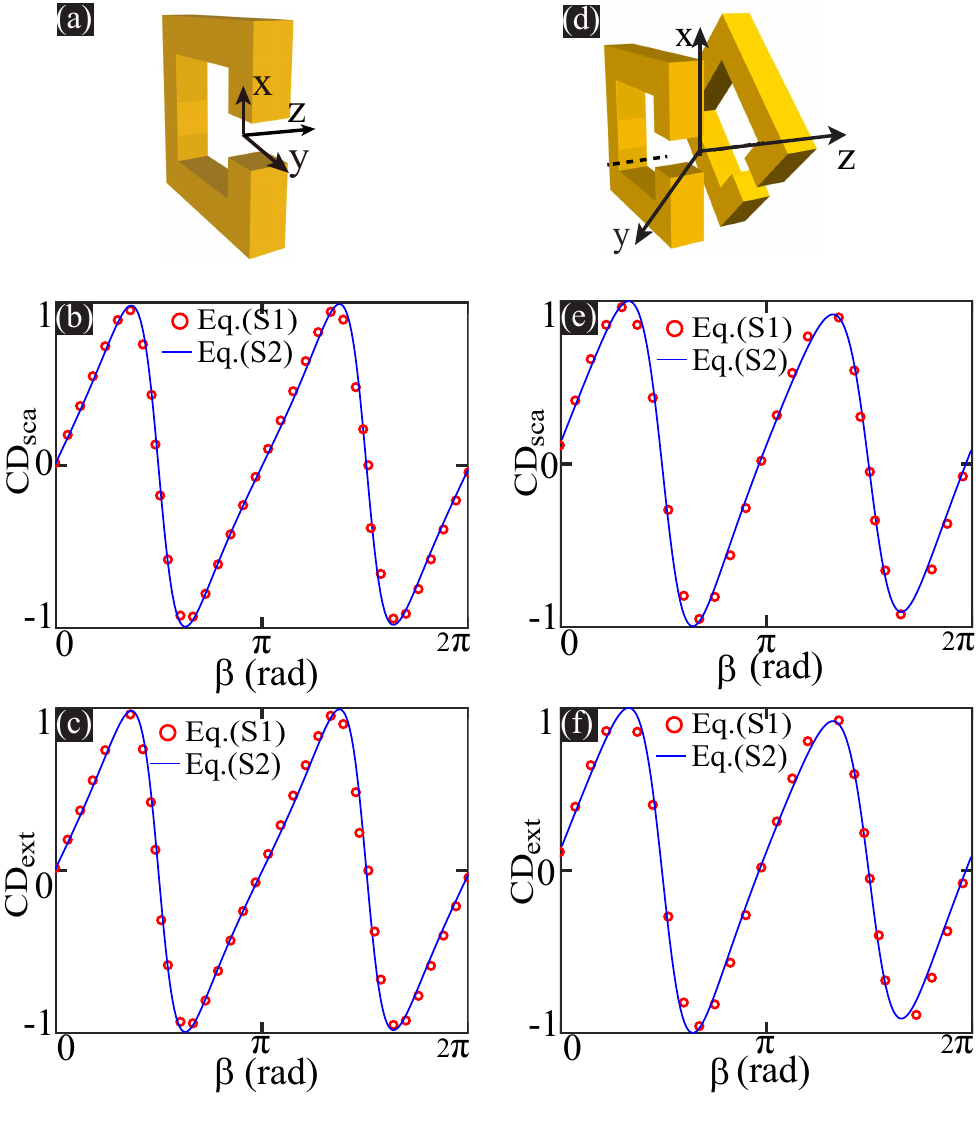}} \caption{\small (a) The achiral SRR and (d) the chiral SRR dimer, which are the same as those studied in Figs. 1 and 2, respectively.   The angular $\rm{CD_{ext,sca}}$ spectra are shown in (b-c) and (e-f), for the structures in (a) and (d), respectively.}
\label{figs1}
\end{figure}

The generalized versions of CD in terms of extinction and scattering cross sections $\rm{C}_{ext,sca}$  are:
\begin{equation}
\label{CD-generalized}
\rm{CD_{ext,sca}}=(\rm {C}_{\rm{ext,sca}}^{\circlearrowright}-\rm {C}_{\rm{ext,sca}}^{\circlearrowleft})/(\rm {C}_{\rm{ext,sca}}^{\circlearrowright}+\rm {C}_{\rm{ext,sca}}^{\circlearrowleft}),
\end{equation}
and it has been clarified in the main text that similar to CD, we have
\begin{equation}
\label{CD-generalized2}
\rm{CD_{ext,sca}}=S_3.
\end{equation}
Here we aim to verify Eq.~(\ref{CD-generalized2}) numerically and show the results in Fig.~\ref{figs1}. The structures we study [Figs.~\ref{figs1}(a) and (d)] are the same as those achiral and chiral scatterers already investigated in Figs. 1 and 2.  The angular $\rm{CD_{ext,sca}}$ spectra are shown in Figs.~\ref{figs1}(b-c) for the achiral scatterer and in Figs.~\ref{figs1}(e-f) for the chiral scatterer.  Other parameters (such as the orientations of the great momentum circles and their parameterization angle $\beta$) for Figs.~\ref{figs1}(b-c) and Figs.~\ref{figs1}(e-f) are the same as those for Fig. 1(f) and Fig. 2(e), respectively. The two sets of results [obtained according to Eq.~(\ref{CD-generalized}) and Eq.~(\ref{CD-generalized2}), respectively] shown for both scenarios agree well, confirming the equivalence between $S_3$ and generalized $\rm{CD_{ext,sca}}$.

%

\section{(\textbf{\uppercase\expandafter{\romannumeral3}}). $S_3$ distributions of  an inversion-symmetric scatterer and its CD responses}

\begin{figure}[htbp]
\centerline{\includegraphics[width=8.5cm]{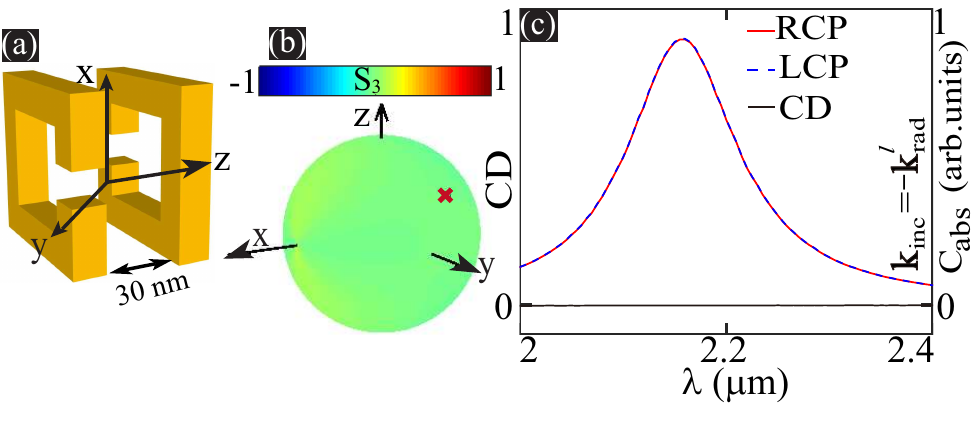}} \caption{\small (a) An inversion-symmetric SRR dimer and (b) the $S_3$ parameter distribution for radiations of the QNM supported at $\tilde{\omega}_{4}=(8.68\times10^{14}-2.599\times10^{13}\rm{i})$ rad/s. A random singularity direction  $\mathbf{k}_{\rm{rad}}^{l} (\theta=58^\circ, \phi=110^\circ)$ is marked ($\times$). (c) CD and absorption spectra for RCP and LCP incident waves are shown with $\mathbf{k}_{\rm{inc}}=-\mathbf{k}_{\rm{rad}}^{l}$.}
\label{figs2}
\end{figure}

Our analyses in the main text show that if an inversion-symmetric structure supports an individual QNM, then the mode is everywhere linearly-polarized ($S_3=0$) on the momentum sphere and thus $\rm{CD}=0$ for all incident directions. This is confirmed numerically in Fig.~\ref{figs2}: the SRR dimer exhibits inversion symmetry [Fig.~\ref{figs2}(a)] and supports dominantly one QNM at eigenfrequency $\tilde{\omega}_{4}=(8.68\times10^{14}-2.599\times10^{13}\rm{i})$ rad/s (the corresponding wavelength for the real part of $\tilde{\omega}_{4}$ is $\lambda=2.17~\mu$m); the $S_3$ distribution of the QNM is presented in Fig.~\ref{figs2}(b), where a random direction is indicated ($\times$). It is clear that $S_3\approx0$ throughout the momentum sphere and thus radiations are linearly polarized along arbitrary directions. Antiparallel to this direction the corresponding CD and absorption spectra are summarized in Fig.~\ref{figs2}(c).

\section{(\textbf{\uppercase\expandafter{\romannumeral4}}). Applications for elliptical dichroism maximization}
\label{berry-dennis model}
In our main text we have discussed only CD for incident circularly-polarized waves of opposite handedness.  Similar investigations can be extended to  elliptical dichroism (ED) for incident waves of orthogonally elliptical polarizations.  Equation~(4) in our main text is reproduced here as follows:
\begin{equation}
\label{expansion-coefficient3}
\alpha(\omega)\propto \tilde{\mathbf{E}}_{\mathrm{rad}} \cdot \mathbf{E}_{\mathrm{inc}}, ~~\mathbf{k}_{\rm{inc}}=-\mathbf{k}_{\rm{rad}},
\end{equation}
which indicates that for generally elliptically-polarized $\tilde{\mathbf{E}}_{\mathrm{rad}}$, there must be two elliptical polarizations which can maximally or not excite the QNM when incident antiparallel to the radiation direction $\mathbf{k}_{\rm{inc}}=-\mathbf{k}_{\rm{rad}}$. Let us assume that the radiation $\tilde{\mathbf{E}}_{\mathrm{rad}}$ is along the direction $\mathbf{k}_{\rm{rad}}$ and characterized by normalized Jones vector (in linear basis of $\mathbf{e}^\theta$-$\mathbf{e}^\phi$, where $\mathbf{e}^\theta$ and $\mathbf{e}^\phi$ are unit vectors in the spherical coordinate system)~\cite{YARIV_2006__Photonics}:  $\mathbf{J}_{\rm{rad}}=(\varsigma,\tau)$, with $|\varsigma|^2+|\tau|^2=1$. For light incident antiparallel to the radiation direction $\mathbf{k}_{\rm{inc}}=-\mathbf{k}_{\rm{rad}}$, according to Eq.~\ref{expansion-coefficient3}: the incident wave that can maximally excite the QNM is characterized by the Jones vector $\mathbf{J}_{\rm{inc}}^\uparrow=(\varsigma^*,\tau^*)$ with normalized excitation coefficient $\alpha^\uparrow=1$, where $^*$ denotes complex conjugate; while the incident wave that cannot excite the QNM corresponds to a Jones vector $\mathbf{J}_{\rm{inc}}^\downarrow=(\tau,-\varsigma)$ with null excitation coefficient $\alpha^\downarrow=0$. We emphasize that all the Jones vectors for both radiations and incident waves are defined in the same linear basis of $\mathbf{e}^\theta$-$\mathbf{e}^\phi$ defined for the radiation direction. The aforementioned two incident waves are obviously orthogonal since $(\mathbf{J}_{\rm{inc}}^\uparrow)^* \cdot \mathbf{J}_{\rm{inc}}^\downarrow=0$. The absorption cross sections associated with two such incident polarizations are:
\begin{equation}
\label{absorption}
\rm{C}_{\rm{abs}}^{\uparrow,\downarrow}  \propto |\alpha^{\uparrow,\downarrow}|^2=1,0
\end{equation}

\begin{figure}[htbp]
\centerline{\includegraphics[width=9cm]{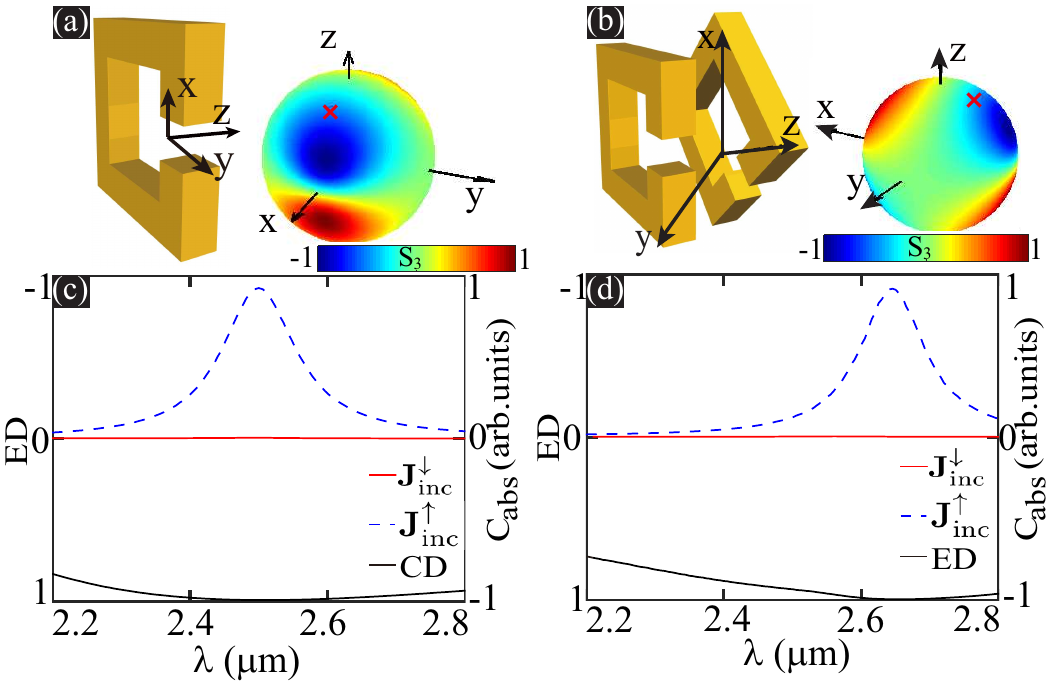}} \caption{\small Structures and mode radiation $S_3$ parameters for achiral and chiral scatterers in (a) and (b), which are respectively reproduced from Figs. 1 and 2. For each scenario a radiation direction is selected along which the radiation is elliptically polarized: $\mathbf{k}_{\rm{rad}}^{\rm{ach}}(\theta=33^\circ, \phi=0^\circ)$ and $\mathbf{k}_{\rm{rad}}^{\rm{ch}} (\theta=21^\circ, \phi=196^\circ)$. ED and absorption spectra for the achiral and chiral cases are shown respectively in (c) and (d).}
\label{figs3}
\end{figure}

The ED is defined as:
\begin{equation}
\label{ED-original}
\rm{ED}=(\rm {C}_{\rm{abs}}^{\uparrow}-\rm {C}_{\rm{abs}}^{\downarrow})/(\rm {C}_{\rm{abs}}^{\uparrow}+\rm {C}_{\rm{abs}}^{\downarrow}),
\end{equation}
which obviously reaches its ideal maximal value $\rm{ED}=1$ for the two incident orthogonal polarizations characterized by Jones vectors of $\mathbf{J}_{\rm{inc}}^{\uparrow,\downarrow}$. This means that our theory can be employed for ED maximization.

To confirm our analyses we revisit the achiral and chiral scatterers already investigated in Figs. 1 and 2, which together with the $S_3$ parameter distributions for the QNMs supported are reproduced respectively in Figs.~\ref{figs3}(a) and (b).  For the achiral scenario in Fig.~\ref{figs3}(a), a radiation direction $\mathbf{k}_{\rm{rad}}^{\rm{ach}}$ is selected and the radiation Jones vector is $\mathbf{J}_{\rm{rad}}^{\rm{ach}}=(0.963,0.269e{^{-1.37i}})$. The corresponding spectra of absorption and ED are demonstrated in Figs.~\ref{figs3}(c), for incident waves of Jones vectors $\mathbf{J}_{\rm{inc}}^{\uparrow,\downarrow}$. While for the chiral scenario in Fig.~\ref{figs3}(b), a radiation direction $\mathbf{k}_{\rm{rad}}^{\rm{ch}}$ is chosen and the radiation Jones vector is $\mathbf{J}_{\rm{rad}}^{\rm{ch}}=(0.969,0.247e{^{-1.51i}})$. The corresponding spectra of absorption and ED are presented in Fig.~\ref{figs3}(d). It is clear that for both scenarios,  ED are ideally maximized close to the resonant frequency.


\end{document}